\documentclass[structabstract]{aa} 
\newcommand{\kms}{$\mathrm{km\, s^{-1}\, }$}
\newcommand\kmps{{\rm\,km\,s^{-1}}}
\newcommand\yr{{\, \rm yr}}
\newcommand\msun{\, \rm M_\odot} 
\newcommand\rsun{\, \rm R_\odot} 
\newcommand\pc{\, \rm pc} 
\newcommand\pcc{{\, \rm pc^{-3}}}
\newcommand\au{\, \rm AU} 

\newcommand{\gap}{\;\rlap{\lower 2.5pt \hbox{$\sim$}}\raise 1.5pt\hbox{$>$}\;}
\newcommand{\lap}{\;\rlap{\lower 2.5pt \hbox{$\sim$}}\raise 1.5pt\hbox{$<$}\;}

\usepackage{natbib}
\usepackage{graphicx}
\usepackage{txfonts}

\begin{document}
   \title{High-velocity stars in the cores of globular clusters: The
     illustrative case of NGC 2808 \thanks{Based on observations collected at the
       European Organization for Astronomical Research in the Southern
       Hemisphere, Chile (083.D-0444).}}

   \author{N. L\"utzgendorf
          \inst{1}
          \and
          A. Gualandris\inst{2,3}
          \and
          M. Kissler-Patig\inst{1}          
          \and
          K. Gebhardt\inst{4}          
          \and
          H. Baumgardt\inst{5}          
          \and
          E. Noyola\inst{6}
          \and
          J. M. D. Kruijssen \inst{2}
          \and
          B. Jalali\inst{7}
		  \and
          P. T. de Zeeuw\inst{1,8}
          \and
          N. Neumayer\inst{1}
          }

   \institute{European Southern Observatory (ESO),
              Karl-Schwarzschild-Stra\ss e 2, D-85748 Garching, Germany\\
              \email{nluetzge@eso.org}
         \and
			 Max-Planck Institut f\"ur Astrophysik,
                         Karl-Schwarzschild-Stra\ss e 1, D-85748
                         Garching, Germany         
         \and
			 Department of Physics and Astronomy,
                         University of Leicester,
                         Leicester, LE1 7RH, United Kingdom
         \and
			 Astronomy Department, University of Texas at Austin, 
			 Austin, TX 78712, USA 
         \and
			 School of Mathematics and Physics, University of Queensland, 
			 Brisbane, QLD 4072, Australia
         \and
             Instituto de Astronomia, Universidad Nacional Autonoma de Mexico (UNAM), 
             A.P. 70-264, 04510 Mexico
             %\email{c.ptolemy@hipparch.uheaven.space} 
         \and
	         1. Physikalisches Institut, Universit\"at zu K\"oln, 
    	     Z\"ulpicher Stra\ss e 77, 50937 K\"oln, Germany
         \and
			 Sterrewacht Leiden, Leiden University, 
			 Postbus 9513, 2300 RA Leiden, The Netherlands}

   \date{Received February 17, 2012; accepted May 15, 2012}

% \abstract{}{}{}{}{} 
% 5 {} token are mandatory
 
  \abstract {We report the detection of five high-velocity stars in
    the core of the globular cluster NGC 2808. The stars lie on the
    red giant branch and show total velocities between 40 and 45
    \kms. For a core velocity dispersion $\sigma_c = 13.4$
      \kms, this corresponds to up to $3.4 \sigma_c$.  These
      velocities are close to the estimated escape velocity ($\sim 50$
      \kms) and suggest an ejection from the core. Two of
    these stars have been confirmed in our recent integral field
    spectroscopy data and we will discuss them in more detail
    here. These two red giants are located at a projected distance of
    $\sim 0.3\pc$ from the center. According to their positions on the
    color magnitude diagram, both stars are cluster members.}
% aims heading (mandatory) 
{We investigate several possible origins for the high velocities of
  the stars and conceivable ejection mechanisms. Since the velocities
  are close to the escape velocity, it is not obvious whether the
  stars are bound or unbound to the cluster.  We therefore consider
  both cases in our analysis.}
% methods heading (mandatory) 
{We perform numerical simulations of three-body dynamical encounters
  between binaries and single stars and compare the resulting velocity
  distributions of escapers with the velocities of our stars. If the
  stars are bound, the encounters must have taken place when the stars
  were still on the main sequence.  We compare the predictions for a
  single dynamical encounter with a compact object with those of a
  sequence of two-body encounters due to relaxation. If the stars are
  unbound, the encounter must have taken place recently, when the
  stars were already in the giant phase.}
% results heading (mandatory) 
{After including binary fractions and black-hole retention fractions,
  projection effects, and detection probabilities from Monte-Carlo
  simulations, we estimate the expected numbers of detections for all
  the different scenarios. Based on these numbers, we conclude that
  the most likely scenario is that the stars are bound and were
  accelerated by a single encounter between a binary of main-sequence
  stars and a $\sim 10\msun$ black hole. Finally, we discuss the
  origin of previously discovered fast stars in globular clusters, and
  conclude that the case of NGC 2808 is most likely a representative
  case for most other detections of fast stars in globular
  clusters. We show that with the present analysis we are able to
  explain high-velocity stars in the clusters M3 and 47 Tucanae with
  simple dynamical encounters.}  {}

\maketitle

%__________________________________________________________________

\section{Introduction}\label{intro}

High-velocity stars in globular clusters have been a puzzle to
astronomers since the beginning of velocity measurements. Especially
finding fast giant stars in the core of globular clusters has drawn
attention. \cite{gunn_1979} found two fast stars which they called
``interlopers" in the globular cluster M3. The stars are located in
the core of the cluster, both at a projected distance of about $20''$
from the center. With radial velocities of $17.0$ \kms and $-22.9$
\kms relative to the cluster, they move with $3.5$ and $4.5$ times the
central velocity dispersion ($\sigma_c = 4.9$ \kms), respectively. Due
to the high systemic velocity of the cluster itself ($\rm V_r \sim
-147$ \kms) the possibility of these stars being field stars is rather
low. \cite{meylan_1991} found a similar case for the globular cluster
47 Tucanae. The stars which they found have radial velocities of
$-36.7$ \kms and $32.4$ \kms in the reference frame of the cluster
which corresponds to $4.0$ and $3.6$ times the core velocity
dispersion $\sigma_c = 9.1$ \kms. Despite the fact that the low
systemic velocity of the cluster does not allow for an accurate
kinematic statement on the membership of these two stars, their
position in the color-magnitude diagram and the high Galactic latitude
of 47 Tucanae ($b=-44^{\circ}$) both argue for membership.
   \begin{figure*}
  \centering \includegraphics[width=\textwidth]{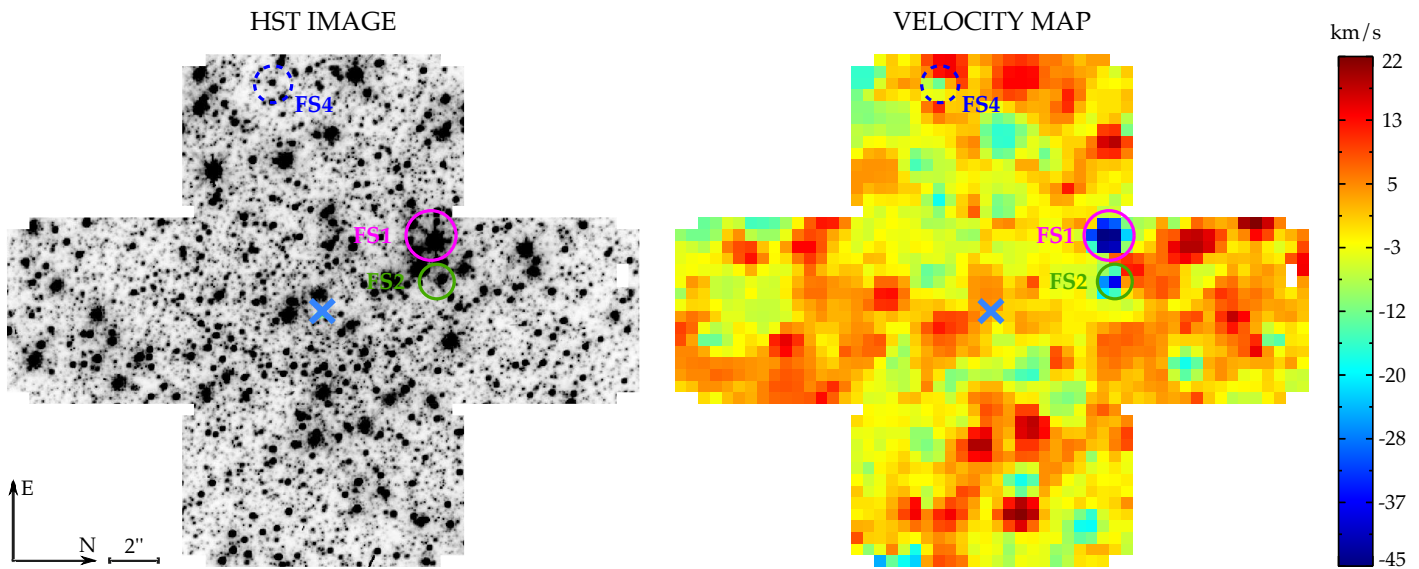}
      \caption{ARGUS field of view (left) and velocity map of NGC
        2808 (right). Visible as bright blue spots are the two
        high-velocity stars FS1 and FS2. Marked with the dashed blue
        line is FS4, the star which does not dominate its spaxel and
        therefore is not detectable. The blue cross marks the center of
        the cluster.}
         \label{vel}
   \end{figure*} 
 
   \begin{figure*}
  \centering \includegraphics[width=\textwidth]{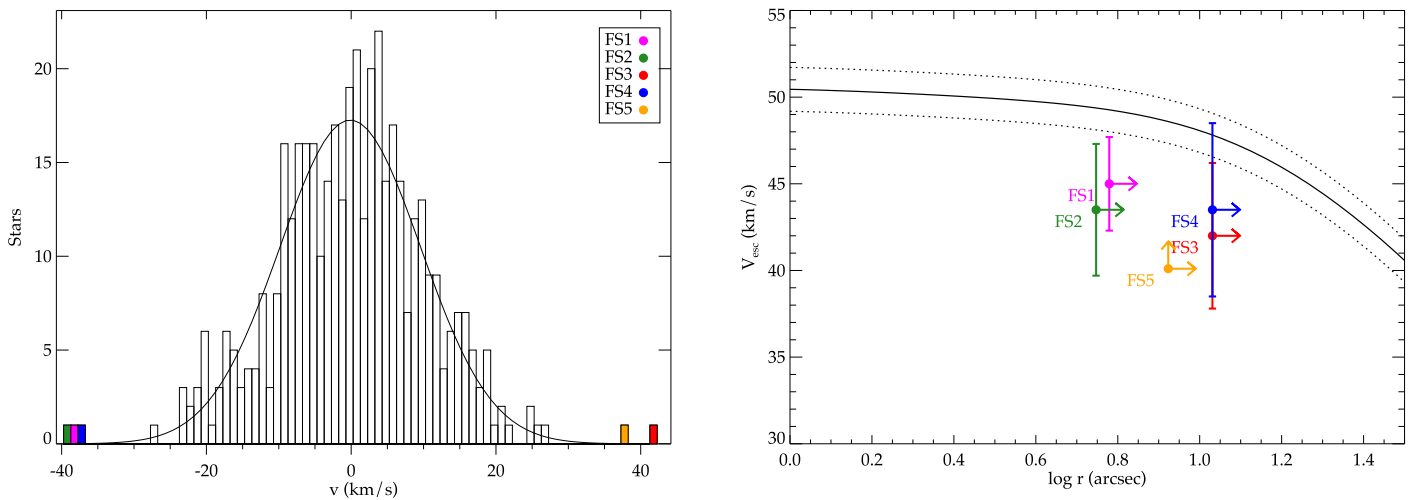}
      \caption{Kinematic properties of the five high-velocity
        stars. Left: The histogram of the Fabry-Perot data in the core
        of NGC 2808. Marked with colors are five high velocity stars
        FS1 to FS5. Right: The escape-velocity profile (solid line)
        and its uncertainties (dotted line) obtained from the density
        profile overplotted by the total velocities and projected
        positions of the five fast stars.}
         \label{hist}
   \end{figure*}

Plausible mechanisms to explain these high-velocity stars are ejection
from the core by recoil from an encounter between a single star and a
binary, between two binary stars, between a binary and an
intermediate-mass black hole (IMBH) at the center or even from encounters of stars with an IMBH binary
\citep[e.g.][]{gualandris_2004, gualandris_2005, baumgardt_2006,
  gvaramadze_2009, mapelli_2005}. It has been recently suggested that globular
clusters may contain central black holes in the mass range of
$10^3-10^4\msun$ \citep{miller_2002,baumgardt_2005} which fall on top
of the relation between the velocity dispersion and black hole mass
seen for galaxies
\citep[e.g.][]{ferrarese_2000,gebhardt_2000}. Kinematic signatures for
intermediate-mass black holes have been found in the globular clusters
$\omega$ Centauri \citep{noyola_2008,noyola_2010}, G1 in M31
\citep{gebhardt_2002,gebhardt_2005} and NGC 6388 \citep{nora11}. High
velocity stars are predicted in globular clusters which host
intermediate-mass black holes. Therefore it is important to study such
stars in more detail when observed in a globular cluster.

The globular cluster NGC 2808 is of high interest in many
regards. From the photometric side, it has a complex extended
horizontal branch \citep{harris_1974, ferraro_1990} which shows
puzzling discontinuities in the stellar distribution along its blue
tail. Furthermore, \cite{piotto_2007} found a triple main sequence
after accurate photometric and proper motion analysis with deep Hubble
Space Telescope (HST) data. This indicates the existence of three
sub-populations with an age of $\sim 12.5$ Gyr and different
metallicities. The kinematic properties of NGC 2808 are also
unusual. By analyzing Fabry-Perot data, \cite{gebhardt_2011} found
rotation in the outer parts of the cluster with a maximum rotation
velocity of $\rm V_{\rm{rot}} \sim 5 $ \kms. The latest radial
velocity measurements of the cluster core by \citet{nora11b} have
shown that NGC 2808 most probably does not host an intermediate-mass
black hole: the best fit model is consistent with no IMBH and gives an
upper limit of $M_{BH} < 1 \times 10^4 \msun$.

In this work we report on the discovery of five high-velocity stars in
NGC 2808 and discuss several explanations and ejection scenarios for
the two stars found in our integral field spectroscopy
observations. In Section \ref{sec:rv} we explain the observations and
the method of obtaining single star velocities with integral field
spectroscopy. Further, we introduce a second data set used for
velocity measurements taken from a Fabry-Perot instrument. At the end
of the first Section we determine the escape velocity of NGC
2808. Section \ref{sec:explain} is dedicated to the discussion of
alternative explanations for the observed high velocities including
foreground stars, binaries and atmospheric active stars. Section
\ref{sec:bound} describes the analysis of the Maxwellian velocity
distribution and the probabilities of these stars resulting from
relaxation while Section \ref{sec:dynenc} describes the dynamical
three-body simulations in order to find the most likely acceleration
scenario for the two fast stars. In Section \ref{sec:others} we
briefly discuss other cases of fast star detections in globular
clusters. Finally, we summarize our results and list our conclusions
in Section \ref{sec:con}.

%__________________________________________________________________

\section{Radial velocities}\label{sec:rv}

The data we use were obtained with the GIRAFFE spectrograph of the
FLAMES (Fiber Large Array Multi Element Spectrograph) instrument at
the Very Large Telescope (VLT) in ARGUS (Large Integral Field
Unit). For more details on the observations and reductions we refer to
\citet{nora11b}. From these data we construct a velocity map which is
shown in Figure \ref{vel} together with the corresponding field of
view from the HST image. The two fast stars are visible as bright
``blue" spots and are labeled with FS1 and FS2 in the maps. To
estimate their velocities, we use the output of a shot noise routine
described in \citet{nora11b}. This routine computes for every spaxel
the number of stars that contribute to the light of the spaxel and
indicates which spaxels are dominated by a single star. With this
information we find the spaxel to which our two (bright) fast stars
contribute with more than 80\% of the light and use the radial
velocities derived for these spaxels as the ones of the stars. These
velocities are corrected for the systemic velocity of the cluster
\citep[$\rm V_r = 93.6 $ \kms ][]{nora11b}.

A second measurement is obtained from the dataset observed by
\cite{gebhardt_2011} with the Rutgers Fabry Perot on the Blanco 4-m
telescope at Cerro Tololo Inter-American Observatory (CTIO). This
dataset contains 3634 stars and covers the center of the cluster as
well as regions up to 4 arcmin radius. We match this dataset with our
photometric catalog from the HST image and identify the two fast
stars. Despite crowding in the inner regions, the two stars are
resolved and can be identified in the Fabry-Perot dataset. In addition
to FS1 and FS2, we find three more stars with velocities values
between $36.0$ \kms and $42.0$ \kms inside the core radius labeled
with FS3, FS4 and FS5. Table \ref{tab_prop} lists the velocity
measurements of all the stars as well as other properties derived from
the HST data. Besides FS1 and FS2 only FS4 is also located inside the
field of view of the combined ARGUS pointing (see Figure
\ref{vel}). However, due to its low contribution of $\sim 55$ \% to
the single spaxel we measure a much lower velocity in the IFU data
than with the Fabry-Perot. The remaining stars lie outside the IFU
pointing. Figure \ref{hist} shows the histogram of the radial
velocities inside the core from the Fabry-Perot data set. The five
fast stars are color-coded and visible as outliers of the
distribution.

In addition to the spectroscopic data sets we also received proper
motions for four out of the five stars (private communication, Jay
Anderson). As listed in Table \ref{tab_prop} the proper motions
indicate that all stars are cluster members and increase the total
velocities up to $40 -45$ \kms. FS3, unfortunately, lies outside the
proper motion dataset and therefore can only be assigned a lower limit
of velocity. Since our measurements are only complete for FS1 and FS2
we decided to limit the following analysis to these two stars.

\begin{figure}
   \centering
   \includegraphics[width=0.48\textwidth]{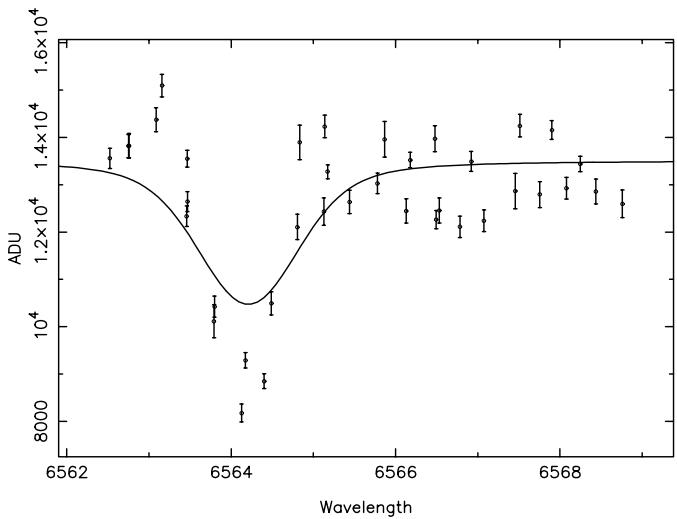}
      \caption{Fabry-Perot photometric points for FS1. Overplotted is the best fit of the model for the H$\alpha$ line.}
         \label{fp}
   \end{figure}

Figure \ref{fp} shows the photometric data points of the Fabry-Perot
for FS1 and the best fit to its spectrum. The data of FS1 is not well
reproduced by the model (solid line). In order to compare velocities
from both datasets the velocities are corrected for the systemic
motion of the cluster by computing the average velocity of the central
region. The results from both data sets show high velocities for these
stars. However, for FS1 the two measurements differ by $7$ \kms,
i.e. more than $1 \sigma$, from each other. The star is close to the
saturation limit and therefore causes difficulties when measuring the
velocity. Also, measuring individual velocities with integrated light
can lead to larger errors since other stars always contribute some
amount of light to the analyzed spectrum. For FS2 the measurements
agree within the error bars. Thus, the difference for FS1 is probably
not due to a systematic shift. The discrepancy for FS1 can have
several reasons: a) the velocity of this star is variable. This
suggests a binary or long period variable and is discussed in more
detail in Section \ref{sec:explain}. b) At least one of the
measurements is affected by a large error. Considering the quality of
the fit of the Fabry-Perot data we consider the spectroscopic
measurement more accurate.

As a next step, we calculate the escape velocity of the cluster as a
function of radius. This value gives the information needed to decide
whether the stars are bound or unbound. We obtain an escape velocity
profile by parametrizing our surface-brightness profile with a
multi-gaussian expansion \citep[MGE,][]{emsellem_1994} as implemented
in the Jeans Anisotropic Models (JAM) by \cite{cappellari_2008}. The
parametrization allows an easy way of deprojecting the profile after
multiplying it with the derived $M/L$ profile from our kinematic fits
\citep[see][]{nora11b}. The gravitational potential generated by the
deprojected density of stars is given by Equation (39) of
\cite{emsellem_1994}:\begin{equation}
\label{eq:pot}
 \Phi(R,z) = - \sqrt{2/\pi} \ G \int\limits_{0}^{1} \sum\limits_{j=1}^{N} \frac{M_j \mathcal{H}_j(u)}{\sigma_j} \mathrm{d} u,
\end{equation}
where $G$ is the gravitational constant, $M_j$ and $\sigma_j$ the
enclosed mass and the dispersion of each of the $N$ Gaussians,
respectively and with

\begin{equation}
\label{eq:hi}
\mathcal{H}_j(u) = \frac{\mathrm{exp}\left\{ \frac{-u^2}{2\sigma_j^2} \left[ R^2 + \frac{z^2}{1-(1-q_j^2)u^2}\right]\right\}}{\sqrt{1-(1-q_j^2)u^2}},
\end{equation} 
a function of the integration variable $u$, the deprojected axial
ratio $0 \le q_j \le 1.$, and the cylindrical coordinates $R$ and $z$.

The escape-velocity profile is then evaluated by calculating the
difference in the potential at a radius of $r$ and the potential at
the tidal radius \citep[$r_t = 35\pc$,][]{harris_1996} as given by the
equation:

\begin{equation}
\label{eq:vesc}
\mathrm{V}_{esc}(r) = \sqrt{2\ (\Phi(r_T) - \Phi(r))}
\end{equation}

The resulting profile is shown in Figure \ref{hist}. Overplotted are
the velocities of the five stars at their projected radii. Since the
radii are projected, they only give a lower limit on the actual
distance from the center. Considering the uncertainties of the stellar 
velocities and of the escape velocity from the cluster,
  we find that star FS4 is consistent with being both bound and unbound.
  The remaining stars appear to be bound,but given the large velocity uncertainties we will also investigate ejection mechanisms in the unbound regime.

\begin{table*}
\caption{Radial velocities and photometric properties of the two fast
  stars FS1 and FS2. The table lists the radial velocity measurements
  from the IFU dataset (V$_{IFU}$) and the velocities measured from
  the Fabry-Perot data (V$_{FP}$). $m_V$ and $m_I$ are the V and I
  magnitudes of the stars and $r_{cen}$ is the projected distance from
  the cluster center.}
\label{tab_prop}      % is used to refer this table in the text
\centering
\begin{tabular}{l c | c c c c c c c c c c c c c c c}
\hline \hline
\noalign{\smallskip}

				&	  	&  \multicolumn{3}{c}{FS1} &	\multicolumn{3}{c}{FS2} &	\multicolumn{3}{c}{FS3}&	\multicolumn{3}{c}{FS4} &	\multicolumn{3}{c}{FS5} 		\\
\noalign{\smallskip}
\hline
\noalign{\smallskip}
$\alpha$   		&		&  \multicolumn{3}{c}{$-09:12:03.620$}     	&   \multicolumn{3}{c}{$09:12:03.346$}	 &	\multicolumn{3}{c}{$09:12:01.783$} &    \multicolumn{3}{c}{$ 09:12:04.746$}	&   \multicolumn{3}{c}{$09:12:03.928$}	    \\
$\delta$   		&	 	& \multicolumn{3}{c}{$-64:51:43.47$}&  \multicolumn{3}{c}{$-64:51:43.16	$}&   \multicolumn{3}{c}{$-64:51:41.64$}&    \multicolumn{3}{c}{$-64:51:50.79$}&   \multicolumn{3}{c}{$-64:51:41.93$}	    \\
$r_{cen}$  		& [pc]	& \multicolumn{3}{c}{$0.28$}&  \multicolumn{3}{c}{$0.26$}&  \multicolumn{3}{c}{$0.50$}&     \multicolumn{3}{c}{$0.50$}&   \multicolumn{3}{c}{$0.39$}		     	\\
$m_V$	   		& [mag]	& $13.56  $&$\pm$&$ 0.01$	&  $14.47 $&$\pm$&$ 0.01$	&   $14.25 $&$\pm$&$ 0.01$	&     $15.02 $&$\pm$&$ 0.01$	&   $15.68 $&$\pm$&$ 0.01$	\\
$m_I$	   		& [mag]	& $11.35 $&$\pm$&$ 0.02$	&  $13.06 $&$\pm$&$ 0.02$	&   $15.51 $&$\pm$&$ 0.02$	&     $16.02 $&$\pm$&$ 0.02$	&   $16.78 $&$\pm$&$ 0.01$	\\
V$_{z,\rm IFU}$	& [\kms]& $-44.0 $&$\pm$&$ 1.8$ 	&  $-37.4 $&$\pm$&$ 0.3$	&   \multicolumn{3}{c}{-}	&     $-12.0 $&$\pm$&$ 6.0$		&   \multicolumn{3}{c}{-}   \\
V$_{z,\rm FP}$ 	& [\kms]& $-38.0 $&$\pm$&$ 2.3$ 	&  $-39.7 $&$\pm$&$ 2.4$	&   $42.0 $&$\pm$&$ 4.2$	&     $-36.0 $&$\pm$&$ 6.3$	 	&   $38.0 $&$\pm$&$ 3.9$	\\
V$_{x,\rm PM}$ 	& [\kms]& $8.6 $&$\pm$&$ 10.3$    	&  $17.8 $&$\pm$&$ 7.0$		&   \multicolumn{3}{c}{-}			&     $22.4 $&$\pm$&$ 1.3$	 	&   $-12.6 $&$\pm$&$ 2.4$	\\
V$_{y,\rm PM}$ 	& [\kms]& $4.0 $ &$\pm$&$ 5.0$    	&  $13.2 $&$\pm$&$ 8.0$		&   \multicolumn{3}{c}{-}			&     $-9.6 $&$\pm$&$ 1.5$	 	&   $2.7 $&$\pm$&$ 1.5$	    \\
\noalign{\smallskip}
\hline
\noalign{\smallskip}
V$_{tot}$  		& [\kms]& $45.0 $&$\pm$&$ 2.7$   	&  $43.5 $&$\pm$&$ 3.8$    	&    \multicolumn{3}{c}{$> 42 $}		&     $43.5 $&$\pm$&$ 5.0$	 	&   $40.1$&$\pm$&$ 0.8$	 \\      
\noalign{\smallskip}
\hline
\end{tabular}
\end{table*}

%__________________________________________________________________

\section{Possible explanations } 
\label{sec:explain}

Despite the uncertainties in the exact velocity, both stars lie more
than $3 \sigma_c$ above the systemic velocity of the cluster, where
$\sigma_c = (13.4 \pm 0.2)\kmps$ \citep{nora11b} represents the
central velocity dispersion. Before investigating dynamical
acceleration scenarios for the stars, we consider possible alternative
explanations for the high velocities.

\begin{figure}
  \centering
  \includegraphics[width=0.5\textwidth]{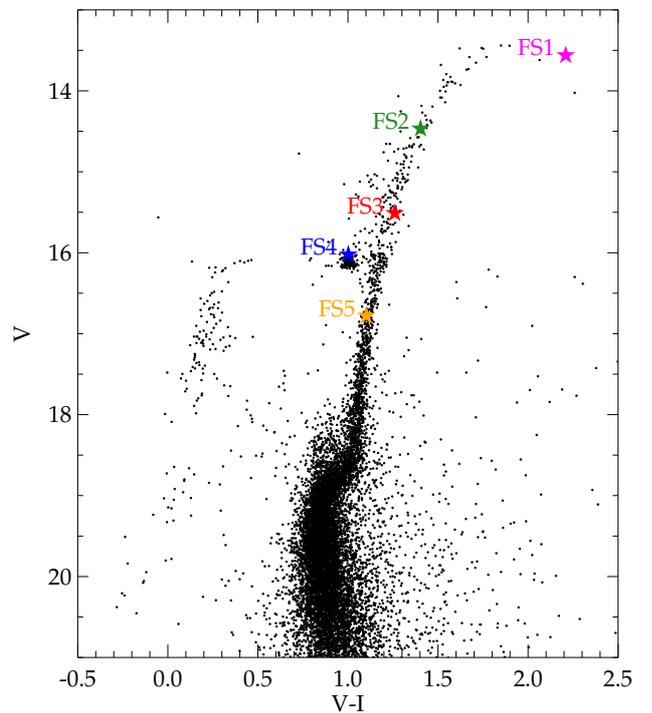}
  \caption{Color-magnitude diagram of NGC 2808 overplotted with
    the positions of the high-velocity stars FS1 to FS4. All of them
    are located on the red giant branch or the horizontal branch and
    therefore most likely cluster members.}
  \label{cmd}
\end{figure}

\subsection{Foreground stars}\label{subsec:foreground}

The first possibility we consider is that the two stars are field
stars and therefore moving with a different velocity than the
cluster. We identify the two stars in our HST image (see Figure
\ref{vel}) and study their position on the color-magnitude diagram
(CMD) which was obtained in \citet{nora11b}. Figure \ref{cmd} shows
the position of all the stars in the CMD. Both stars are located on
the giant branch, which suggests cluster membership. The three other
stars are also most likely cluster members. FS1 is the brightest star
in our pointing and sits at the top of the giant branch. The fact that
the brightest star in our dataset is also the fastest is
suspicious. This star is most likely a long period variable star and
will be discussed in more detail in Section \ref{subsec:var}. The
second star (FS2), however, lies in the central part of the giant
branch about one magnitude fainter than FS1 and does not seem to show
any peculiar photometric properties.

The fact that both stars lie on the giant branch of the cluster does
not strictly exclude them from being foreground stars. Stars at a
distance of a few hundred pc may also appear superposed on the giant
branch of NGC 2808. In the case of FS2, a lower main sequence
foreground star would be projected up on the giant branch if its
apparent magnitude was increased by 8 to 9 magnitudes. This would
imply that the star is located 40 - 60 times closer than a giant of
similar apparent magnitude. That is, at a distance of $240-150\pc$ from
us. FS1 is even more extreme. With a color $(B-V) \sim 1.9$, an
equivalent dwarf star would have an apparent brightness increased by
more than 12 magnitudes. This would mean that the star is 240 times
closer to the sun than its giant equivalent and therefore at $40\pc$
distance from us.

In general the probability of contamination through galactic field
stars is rather low. According to a galactic model by
\cite{ratnatunga_1985}, the estimated number of field stars per square
arcminute projected onto the central regions of NGC 2808 is $3.4
\times 10^{-2}$ for $0.8 < (B-V) < 1.3$ (FS2) and $1.8 \times 10^{-3}$
for $(B-V) > 1.3$ (FS1). The stellar population synthesis model of the
Milky Way by \cite{robin_2004} can also be used to predict the number
of foreground stars expected in this part of the sky. For a field of
view of $0.5 \, \rm{arcmin}^2$ at the position of $l=282.19^{\circ},
\, b=-11.25^{\circ} $, the model does not predict any stars in the
observed magnitude ranges ($m_V = 13.5 - 14.5$ mag, $m_I = 11.5 -
13.0$ mag). Also, the high radial velocity of NGC 2808 makes it easier
to distinguish between field stars and cluster members.

Finally, neither star shows peculiar motions in the plane of the sky
and the tangential velocities fall inside the velocity distribution of
the cluster (see Table \ref{tab_prop}). This strongly argues for
membership and a dynamical connection between the stars and
cluster. Considering all these facts, we conclude that both stars are
giants and associated to the cluster.

\subsection{Binaries}\label{subsec:bin}

The binary fraction in typical globular clusters is about 10 \%
\cite[e.g.][]{rubenstein_1997,davis_2008,dalessandro_2011}. Due to
their higher dynamical mass, binaries tend to sink to the cluster
center and become more concentrated in that region. This is why it is
likely that binaries are present in our field of view. The possibility
that these two fast stars are binaries, however, is rather low.

The fact that we measure very similar radial velocities at different
epochs (IFU and Fabry-Perot datasets) with a time difference of 15
years already speaks against a binary scenario. Within this time
period we should have seen large differences between the two
measurements if the velocities were periodical. If we assume that the
radial velocities that we measure for our two stars are their orbital
velocities, a $\sim 1\msun$ companion would imply a separation of
$\sim 0.5\au$.  The two stars are located on the giant branch and
their radii are of about $7\rsun$ (FS1) and $70\rsun$ (FS2) (most
likely lower limits), which are comparable to the presumed
separation. This would indicate that these systems are in deep contact
and that we should see signatures of accretion and mass transfer in
the form of X-ray and UV excesses. \cite{servillat_2008} obtained
X-ray observations with Chandra for the core of NGC 2808. Comparing
their observations with our image, we see no match between our two
stars and any of the X-ray sources detected. We also investigate a
near and a far UV image obtained with the Space Telescope Imaging
Spectrograph (STIS) on board the Hubble Space Telescope (Program ID:
8511). The stars are visible in the near UV but no source was detected
in the far UV image. We conclude that these stars show no excess in
either X-ray and UV which excludes almost all scenarios of contact
binaries.

\cite{sommariva_2009} investigate the binary fraction of the globular
cluster M4 via spectroscopy. They find 57 binary star candidates out
of 2469 observed stars with 4 candidates inside the core
radius. Looking at their table of candidates we find 7 binaries which
show velocities $ > 20$ \kms in one of their orbital phases. Since the
velocity dispersion of NGC 2808 is over twice as high as the one of M4
($\sim 4$ \kms), total velocities around 40 \kms would be possible if
the binary itself approached with a velocity $> 20$ \kms. However, all
high velocity binaries in M4 lie in the upper main sequence and the
turn-off region. There is no high velocity binary observed on the
giant branch. To make a quantitative statement, we estimated the
expected fraction of binaries with these properties and velocities
higher than 45 \kms with binary population synthesis.

We assume a binary fraction of 10~\% and use a log-normal orbital
period distribution with a mean of $\langle$log (P/days)$\rangle \ =
4.8$ and a sigma of $2.3$ as obtained by \cite{duquennoy_1991}. The
binary components are chosen randomly from an evolved (over $12$ Gyr)
Kroupa initial mass function (IMF) to account for the age of NGC
2808. For the eccentricities we assume a thermal distribution of $f(e)
= 2e$. Thus, for every binary we compute $M_1, M_2, P$ and $e$. From
$P$ we derive the semi-major axis $a$ and place both components ($M_1,
M_2$) at random positions in their orbits. We then distribute the
binary orientation randomly in space and add systemic velocities drawn
from a velocity distribution with $\sigma = 13.4$ \kms. Finally we
observe the binaries from a random position and count how often a
giant star or main-sequence star with $M > 0.8 \msun$ has velocities
$|$V$| >$ 45 \kms. As additional boundary conditions we required that
the pericenter distance of the binary is larger than the sum of the
stellar radii, i.e. $a(1 − e) > (R_{\ast 1} +R_{\ast 2})$

Among the 1000 bright stars in our pointing we expect 0.0054 stars
like FS2 ($R_{\ast} \sim 7 \rsun$) and only 0.00012 stars like FS1
($R_{\ast} \sim 70 \rsun$) if their velocities originate from orbital
binary velocities. To summarize, it is possible but highly unlikely
that the high velocities we observe are due to orbital velocities in
binary systems.

\begin{figure}
  \centering
  \includegraphics[width=0.48\textwidth]{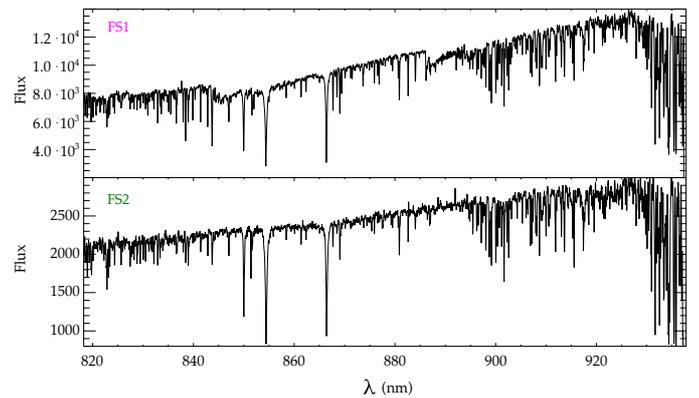}
  \caption{Spectra of the two fast stars extracted from the ARGUS
    integral field unit in the Calcium Triplet region. The lines of
    both stars are narrow and confirm the assumption that both stars
    are giants.}
  \label{spec}
\end{figure}

\subsection{Atmospheric active stars} \label{subsec:var}

Another possibility is that these stars have active atmospheres, such
as strong stellar winds or expanding shells which we would see as
approaching velocities. According to its location in the CMD, FS1 is
most likely a Mira star. Studies have shown that the amplitudes of
Mira pulsations do not exceed 30 \kms
\citep[e.g.][]{hinkle_1978,hinkle_1982, hinkle_1997, lebzelter_2000,
  lebzelter_2005, lebzelter_2011}. Our stars show velocities faster
than this limit. However, an explanation could be that these stars
have radial velocities of about 20 \kms on top of their Mira
pulsations.

\cite{lebzelter_2011} recently investigated the long period variables
of NGC 2808.  Unfortunately, they had to exclude the inner regions of
the cluster due to crowding. We identify their long period variable
stars in our pointing and found two matches. The first one is LW 15
with a period of 332 days while the second one, LW 7, does not have a
precise position and is therefore an uncertain candidate with a period
of 21 days. We use the matches in our pointing to compare these
spectra with our two fast stars.  The spectra of the fast stars are
shown in Figure \ref{spec}. No conspicuous features were found in the
Mira spectra nor in the spectra of the fast stars. The lack of studies
of Mira spectra in the optical range, especially in the Calcium
Triplet region, prevents us from making comparisons with other
measurements.

Another argument against the pulsation velocities is the fact that
optical lines usually form in the upper atmospheres where no shock
waves can be observed. Also, these optical velocities are nearly
always directed inward relative to the stellar center-of-mass as found
by \cite{wood_1979}.

It is worth mentioning that we found no asymmetry in any of the lines
used to measure the radial velocity. Assuming that the high velocity
would come from an atmospheric effect, such as winds or expanding
shells, one would expect the lines to be either asymmetric like
p-Cygni profiles or even split into two parts which is often observed
in Mira variables \citep[eg.][]{scholz_1992, lamers_1999}.

We conclude that thermal pulsations are an explanation for the high
velocities of our two stars seem rather unlikely. The range of
velocities reached by these pulsations is not high enough to explain
the peculiar motions of the stars and unlikely to be strongly
represented in optical lines.

\section{The bound case - Maxwellian distribution}
\label{sec:bound}

Having established that the most likely scenario is that the two stars
do indeed have velocities exceeding more than three times the velocity
dispersion of the cluster, we investigate possible acceleration
mechanisms of dynamical nature.  We first consider the case in which
we assume that the stars are bound to the cluster.  Under this
assumption, the stars could have been accelerated either through
encounters as giants or in their main-sequence stage. For the latter
the acceleration can occur through either many two-body encounters or
a single encounter with a more massive object. In this Section we
discuss the first scenario and the likelihood of observing such an
event.

\subsection{Acceleration through uncorrelated two-body encounters} 
\label{subsec:mw}

\begin{figure*}
  \centering
   \includegraphics[width=\textwidth]{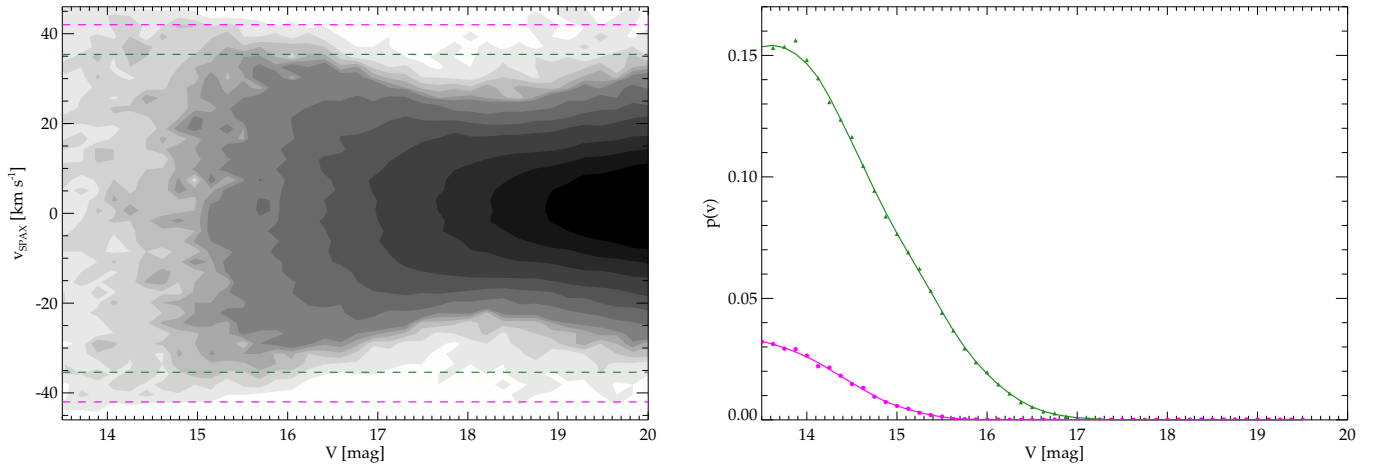}
   \caption{Monte Carlo simulations of recovering the radial
     velocities of v$= 44$ \kms and v$ = 37$ \kms as a function of the
     magnitude of the star. The left panel shows the measured
     velocities of every spaxel from the $10^6$ realizations as a
     density contour plot. The magenta and green lines mark the
     detection limits of $42$ \kms and $35$ \kms for FS1 and FS2
     respectively. A spaxel with a velocity lower than $2$ \kms (which
     corresponds to the error bars) minus the actual radial velocity
     is not considered to be a detection anymore. The right panel
     visualizes the probability of detecting a star with a radial
     velocity of $44$ \kms (magenta) and $37$ \kms (green) as well as
     their smoothed curves.}
   \label{mc}
\end{figure*}

The first possibility we consider is that the stars acquired their
high velocities through relaxation, i.e. a sequence of uncorrelated,
distant encounters with other cluster stars over long timescales. In
such a case they would be bound to the cluster and belong to the tail
of the three-dimensional Maxwellian velocity distribution which is
established by relaxation. However, stars with velocities $\sim 3.5
\sigma$ are quite rare. The chance of detection decreases even more if
one also considers the probability of detecting the high-velocity star
in an integral-field unit (IFU) since the star has to be very bright to be
detected. The final probability is composed of the probability of a
Maxwellian velocity distribution with a certain velocity dispersion
producing a high velocity star, the number of stars in our pointing
and the probability of detecting such a star in our pointing.

In order to make a quantitative statement we calculate these
probabilities in the following way. We use the velocity dispersion
profile obtained in \citet{nora11b} to estimate $\sigma$ at the
position of the two fast stars. From this, we find that the stars have
velocities of $3.2$ and $3.1$ times their local velocity
dispersion. The probability of finding a star with a total velocity V
or higher from a three-dimensional Maxwellian velocity distribution is
then given by:

\begin{equation}
\label{eq:fraction}
 p(x)=1 - \left[{\rm erf}\left(\frac{x}{\sqrt{2}}\right)-\sqrt{\frac{2}{\pi}}x{\rm e}^{-x^2/2}\right],
\end{equation} with $x = \rm V / \sigma$.

With this information we can calculate how many stars we would expect
among all the stars in the pointing using the stars which have been
found in the HST image. In order to obtain the detection probability
we run Monte Carlo simulations for $10^6$ spaxels. For each simulated
spaxel we draw 10 stars from the luminosity function. For 9 of them,
we assign random total velocities from a Maxwellian velocity
distribution with $\sigma = 13.4$ \kms as well as a random spatial
orientation. The tenth star is always assigned a velocity value of
$44$ \kms. We build the spaxel in the following way: for each star we
take the template spectrum, scale it with the flux of the star, and
shift it with the amount of its projected radial velocity. Then we
combine all 10 spectra to a single spaxel spectrum. The velocity of
each spaxel is measured using the penalized pixel-fitting (pPXF)
program developed by \cite{cappellari_2004} as described in
\citet{nora11b}. Figure \ref{mc} shows the result of the Monte Carlo
simulations. Plotted are all the $10^6$ measured spaxel velocities as
a function of the magnitude of the star with V $= 44$ \kms in the form of a
density contour. This shows that the probability of detecting a high
velocity star strongly decreases with the magnitude of the star.

In order to calculate the detection probability, we set the detection
limit equal to the minimal velocities within the error bars (2 \kms),
$42$ \kms for FS1 and $35$ \kms for FS2. Spaxel with velocities below
this limit are not considered to be a detection anymore. The
probability is obtained by deriving the fraction of stars which are
faster than these limits in overlapping bins of 100 stars each. The
right panel of Figure \ref{mc} shows the smoothed curve of the
probability for FS1 (magenta) and FS2 (green). The plot shows that the
probability to detect a star with a radial velocity of $37$ \kms in
the IFU is higher by almost a factor of 5 than a star with $44$ \kms
in the radial component and that for stars fainter than $m_V = 17$
both velocities are not detectable anymore.

To derive the expected number of stars with this configuration we
multiply the luminosity function of stars in our pointing with the
probability from the Maxwellian distribution and the detection
probability from the Monte Carlos simulations and integrate over all
magnitudes for both stars. As a final result we derive:

\begin{eqnarray}
\label{eq:mw}
N_{r,\rm FS1} & = & 0.02 \ \rm{stars} \\
N_{r,\rm FS2} & = & 0.30 \ \rm{stars} 
\end{eqnarray}

For further analysis we also derive the pure detection probability $f_b$, which accounts for (1) the geometric probability that the velocity component along the line of sight exceeds a certain threshold velocity, and (2) the probability that the star is massive and bright enough to be detected in the IFU. The probabilities are derived in the same way as described above but
without multiplying by the probability from the Maxwellian
distribution. Thus we integrate only the probability function from the
Monte Carlo simulations multiplied by the number of stars over the
entire magnitude range. The derived detection probabilities for stars
like FS1 and FS2 are $f_{b,\rm FS1} = 4 \times 10^{-4}$ and $f_{b,\rm
  FS2} = 4 \times 10^{-3}$, respectively.

We conclude that neither FS1 nor FS2 is likely to originate from the
Maxwellian velocity distribution. The probability for a star such as
FS2 is higher than that for FS1 by an order of magnitude but still too
low to explain the observations with this mechanism. It is therefore
unlikely that FS1 and FS2 acquired their high velocities through
relaxation and we study other acceleration scenarios in the following
Sections.

\section{Dynamical encounters}
\label{sec:dynenc}

The numbers we derive from the analysis of the Maxwellian distribution
and observational effects are too small to explain the observed
stars. Therefore we test for alternative acceleration scenarios due to
dynamical three-body encounters of two different types of stars
(main-sequence stars and giants) as described in this Section.

\subsection{Acceleration in the main-sequence stage}
\label{subsec:msenc}

Here we study a scenario in which FS1 and/or FS2 were accelerated by a
single dynamical encounter at some time during the main-sequence
stage. To test this possibility we perform numerical three-body
scattering experiments \citep[e.g.][]{HF1980, HB1983, SP1993, HHM1996,
  gvaramadze_2008, gvaramadze_2009, gvaramadze_2011} involving
main-sequence stellar binaries and single stars/compact objects of
different kinds. The simulations are carried out with the {\tt sigma3}
package included in the STARLAB software environment \citep{MH1996}.

\begin{figure*}
  \centering \includegraphics[angle=270,width=0.49\textwidth]{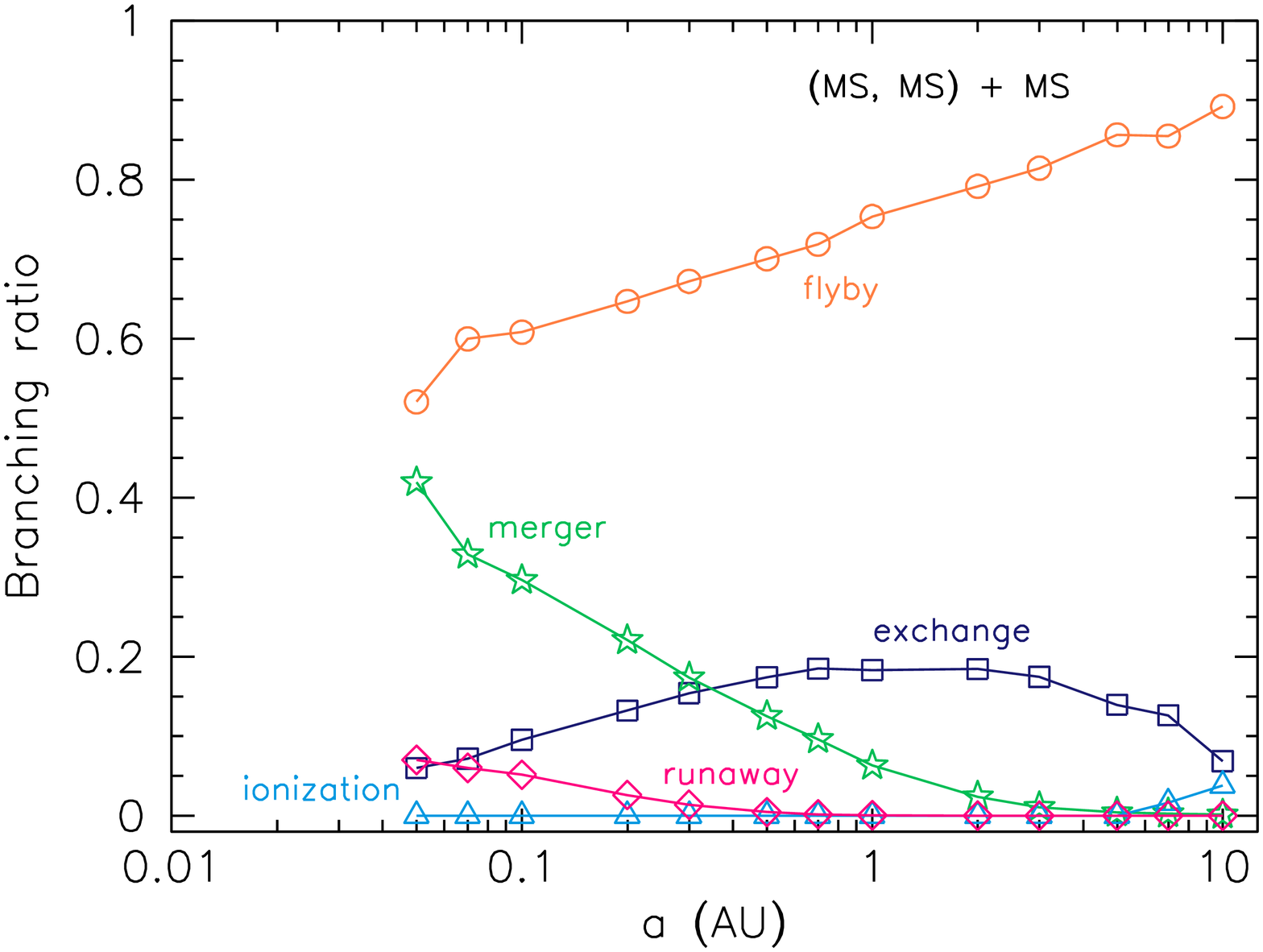}
  \centering \includegraphics[angle=270,width=0.49\textwidth]{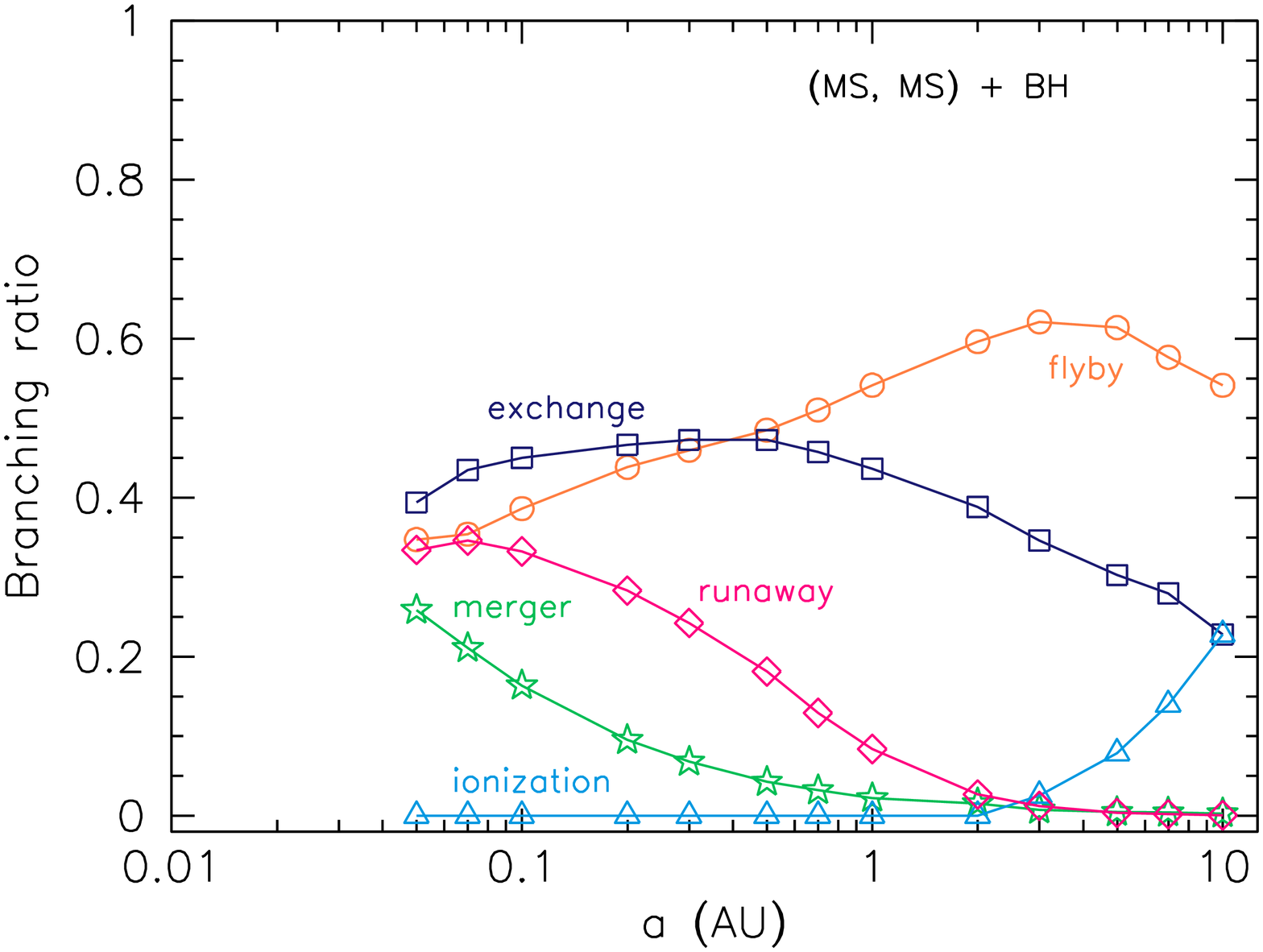}
  \caption{Branching ratios versus initial binary separation for the
    outcomes of encounters between a stellar binary and a
    main-sequence star E1 (left) or a stellar mass black hole E4
    (right). The label ``runaway'' refers to encounters that produce
    an escaper with velocity larger than $45\kmps$.}
  \label{fig:br}
\end{figure*}

We adopt a mass of $m = 0.8\msun$ and a radius of $R =
0.8\,(M/\msun)^{0.7}\rsun$ \citep{hh81} for both stars, and we set the
characteristic ejection velocity to $V_{\rm ej} = 45\kmps$.
\begin{table}
\caption{Dynamical encounters between a main-sequence stellar binary
  and a single star. Here MS indicates a main-sequence star, WD a
  white dwarf, NS a neutron star and BH a stellar mass black
  hole. Outcomes of interest for the ejection of one of the MS stars
  include flybys (F), exchanges (E) and ionizations (I).}
\label{tab:encms}
\centering     
\begin{tabular}{lll} 
\hline
\noalign{\smallskip}
Name & Encounter & Outcomes   \\
\noalign{\smallskip}
\hline                        
\noalign{\smallskip}
E1 & (MS, MS) + MS  & F, E, I \\
E2 & (MS, MS) + WD  & E, I    \\
E3 & (MS, MS) + NS  & E, I    \\
E4 & (MS, MS) + BH  & E, I    \\
E5 & (BH, MS) + MS  & F, E, I \\
\noalign{\smallskip}
\hline                             
\end{tabular}
\end{table}
We consider four types of binary-single star encounters, which are
detailed in Table \ref{tab:encms}: (1) encounters with a main-sequence
star of the same mass, (2) encounters with a $0.8\msun$ white-dwarf,
(3) encounters with a $1.4\msun$ neutron star and (4) encounters with
a $10\msun$ black hole. We consider a fifth type of encounters,
between a binary containing a main-sequence star and a stellar mass
black hole and single stars. Such binaries originate from the
encounters of stellar binaries with single black holes.

We first discuss the simulations involving stellar binaries.  As is
typical in three-body scatterings, we use a Monte Carlo approach to
randomly generate the angles that define the spatial orientation of
the binary with respect to the incoming star. The binary eccentricity
is randomly drawn from a thermal distribution \citep{heggie1975},
having set a maximum value to guarantee that the binary components do
not come into contact at the first pericenter passage.  The impact
parameter $b$ is randomized according to an equal probability
distribution for $b^2$ with a maximum value that is determined
automatically for each set of experiments \citep[see][for a detailed
  description]{gualandris_2004}. The relative velocity between the
single star and the binary center of mass is set equal to the cluster
velocity dispersion. We vary the initial semi-major axis of the binary
from a minimum value $a_{\rm min} = 0.05\au$ set by the radii of the
stars to a maximum value $a_{\rm max} = 10\au$.

While stars are treated as point particles in the integrations, all
inter-particle distances are monitored to identify physical collisions
when the distance between any two stars becomes equal to the sum of
their radii. Since we are interested in the ejection of stars with
high velocity, we neglect all encounters that result in a physical
collision. While it is a priori possible to eject collision
  products during dynamical encounters, the fact that stars FS1 and
  FS2 lie precisely on the color-magnitude diagram of the cluster
  strongly argues against such a scenario.

Possible outcomes of three-body encounters include: flybys of the
single star past the binary (F), exchanges of the incoming star into
the binary (E), with ejection of one of the binary components,
ionization (I) of the binary where three single stars get ejected, and
collisions. For each type of encounter listed in Table
\ref{tab:encms}, we consider all outcomes resulting in the ejection of
(at least) one main-sequence star, and record the ejection
velocity. Ionizations are rare and, for our set of parameters, only
occur in the case of encounters with a black hole. In all the other
cases, the binaries are too hard to be broken apart by the incoming
star (see Figure \ref{fig:br}). In the case of encounters between
equal mass stars, the binary is always too hard to be unbound by the
incoming star.

Figure \ref{fig:br} gives the branching ratios, i.e. the fraction of
all encounters resulting in a particular outcome, as a function of the
initial binary separation, for model E1 and E4.  The relative
importance of flybys, exchanges and mergers is a strong function of
the binary semi-major axis, with mergers occurring in about 40\% of
the cases for the tightest binaries.

\begin{figure}
  \centering \includegraphics[angle=270,width=0.5\textwidth]{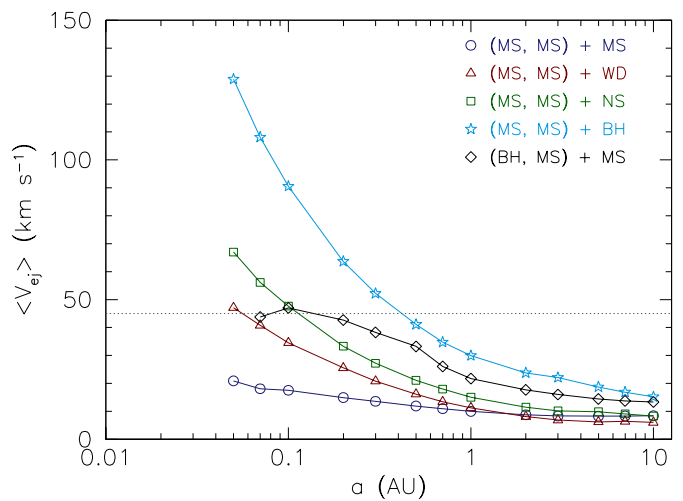}
  \caption{Average ejection velocities of all main-sequence escapers
    for the different types of encounters considered. For model E5, we
    adopt a black hole mass of $30\msun$. The dotted
    horizontal line marks the characteristic velocity of the
    runaways.}
  \label{fig:velo2}
\end{figure}

The average velocity of all main-sequence escapers is shown in
Fig.\,\ref{fig:velo2} for all types of encounters. Encounters with a
main-sequence star produce average recoil velocities significantly
lower than required for the observed runaways, for all considered
values of the binary semi-major axis. Encounters with white dwarfs and
neutron stars produce only slightly larger velocities, and reach the
required $45\kmps$ only for the tightest binaries.  Encounters with a
stellar mass black hole, on the other hand, result in typical ejection
velocities in excess of $45\kmps$ for all values of $a \lap
0.4\au$. Given that these are average velocities, large velocities can
occasionally be achieved even for larger separations.

\begin{figure}
  \centering \includegraphics[angle=270,width=0.47\textwidth]{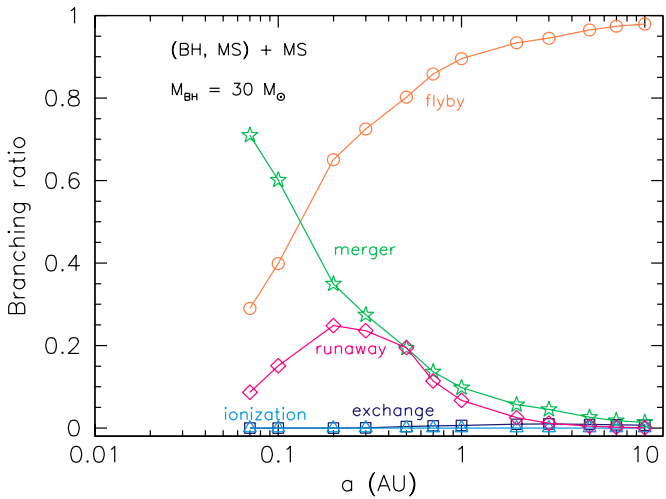}
  \caption{Branching ratios versus initial binary separation for the
    outcomes of encounters of type E5, in the case of a black hole of
    $30\msun$.}
  \label{fig:br2}
\end{figure}

If encounters with stellar-mass black holes lead to exchanges, they are expected to produce a star-black hole binary that will then interact with the dominant population of single, main-sequence stars.
We find that such binaries typically have large eccentricities. For
this reason, when simulating encounters of type E5, we generate
binaries with a suprathermal eccentricity distribution.
The branching ratios for this process are shown in Figure \ref{fig:br2}.
Most encounters of this kind result in simple flybys or mergers. The
average velocity of escaping main-sequence stars is given in
Fig.\,\ref{fig:velo2} for the case of a $30\msun$ black hole. Only
binaries with separation $\lesssim 0.1\au$ produce average velocities
larger than $45\kmps$. On average, higher black holes masses do not result in larger ejection velocities, because a larger fraction of encounters results in a collision.

The rate of ejection of runaway stars like FS1 and FS2 by encounters
with a compact object can be estimated as
\begin{equation}
\mathcal{R} = n\, \Sigma\,  \rm V\,,
\end{equation} 
where $n$ is number density of stars in the core, $\Sigma$ is the
cross-section for the process and V is the relative velocity between
the binary and the black hole. The cross section can be derived from
the scattering experiments as
\begin{equation}
\label{eq:r}
\Sigma = \pi \,b_{\rm max}^2 \, f_{r} \,,
\end{equation} 
where $b_{\rm max}$ represents the maximum impact parameter for the
process under consideration and $f_r$ is the fraction of encounters
resulting in escapers with velocity larger than $45\kmps$ (see
Fig.\,\ref{fig:branch}). The rate for interactions is therefore
\begin{equation}
\mathcal{R} \approx 1.69 \times 10^{-9} \frac{1}{\yr} \,
\left(\frac{n}{10^5 \pcc}\right) \left(\frac{b_{\rm max}}{10
  \au}\right)^2 \left(\frac{\rm V}{10\kmps}\right)\,.
\end{equation}

\begin{figure}
  \centering \includegraphics[width=0.50\textwidth]{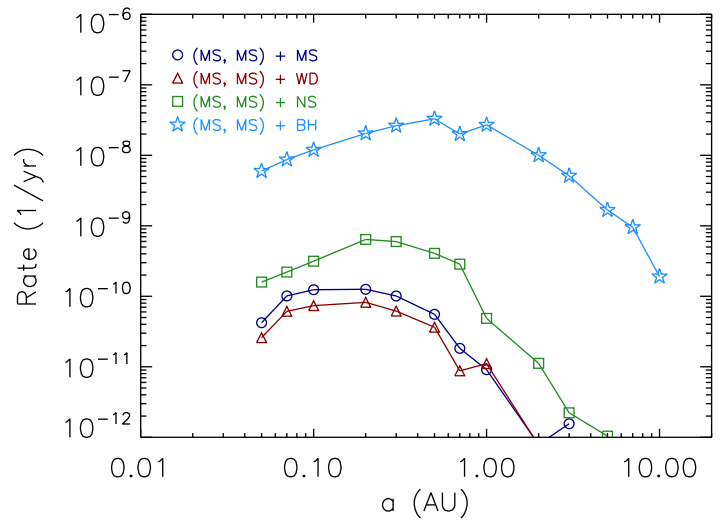}
  \caption{Event rates of the different encounters as a function
    of the separation multiplied by the distribution function of the
    semi-major axis. From this the final event rate can be achieved by
    integrating over the values for the semi-major axis.}
  \label{fig:rate}
\end{figure}

Finally, we multiply the event rate $\mathcal{R}$ as a function of the
binary semi-major axis by the distribution function of the semi-major
axis taken from \cite{duquennoy_1991}. The final rates are shown in
Figure \ref{fig:rate}, where we assume a relative velocity equal to
the dispersion velocity in the core $V = 13\kmps$, and a stellar
density $n = 10^5\pcc$. We obtain the central density by extrapolating
our density profile ($n(0) = 10^6 \pcc$) and multiplying this by a
binary fraction of $10 \%$. We caution that this is a mere working
estimate derived from the total estimated stellar mass and the
structural parameters of the cluster, and $\mathcal{R}$ scales
linearly with the stellar density. The final rate is derived by
integrating over all $a$.

The number of events producing runaway stars with velocity $\gap
45\kmps$ is then

\begin{equation} 
N_r = \mathcal{R} \, N_{\rm BH} \, T_{\rm rlx}\, f_{b}, 
\end{equation}  
where $N_{\rm BH}$ is the number of BHs to be found in the core,
$T_{\rm rlx}$ is the relaxation time of the core (the high velocity of
the stars should decrease due to two-body relaxation within this time
scale) and $f_b$ the detection probability as obtained in Section
\ref{subsec:mw}.

Estimating the number of black holes in the cluster core is not straightforward. There are two factors which play a major role. The first one is the black hole retention fraction $f_{\rm{ret}}$, i.e. the fraction of black holes that does not escape the cluster due to kick velocities acquired during their formation. The second factor is how efficiently equal-mass black holes eject themselves from the cluster core by building a subcore through mass segregation. This process would leave only a few (0-3) black holes in the cluster \citep[e.g.][]{sigurdsson_1993,kulkarni_1993,miller_2002,oleary_2006}. However, \cite{mackey_2007} find different results. In their N-body simulations, larger numbers of black holes are retained because the ejection rate of black holes decreases as the subcluster evolves. In order to estimate an appropriate number of black holes in the core of the globular cluster we use results from our own N-body simulations which are described in L\"utzgendorf et al. (2012, in preparation). Our results show that for a model with a retention fraction of $f_{\rm{ret}} = 0.3$ and $N=128~000$ particles there are only one or two black holes remaining after 12 Gyr. For higher retention fractions such as $f_{\rm{ret}} = 1.0$ this number increases to 20 black holes after 12 Gyr. With $N=50~000$ stars remaining, these values correspond to black-hole fractions of $f_{BH} = N_{BH}/N_{\ast} = 2 \times 10^{-5}$ and $4 \times 10^{-4}$ respectively, equivalent to 20 to 500 black holes in a cluster with $1 \times 10^6$ stars. These values are lower limits though, since we are not only interested in the number of black holes at the current time, but more so in the average number of black holes over the entire lifetime of the cluster. Therefore, an average black hole number of $N_{BH} \sim 100$ seems adequate as a first-order estimate.

Adopting a relaxation time $T_{\rm rlx} = 10^8\yr$
\citep{harris_1996}, we find the expected number of runaway stars
produced by encounters between a main-sequence binary and a stellar
mass black hole to be:

\begin{eqnarray}
\label{eq:msenc}
N_{r,\rm FS1} & = & 0.3 \ \rm{stars} \\ 
N_{r,\rm FS2} & = & 3.4 \ \rm{stars}
\end{eqnarray}

The fact that the expected number of stars like FS1 is an order of
magnitude lower than that of FS2 comes from the detection
probabilities. A star like FS1 is less likely to be observed due to
its brightness and the fact that its velocity vector has almost no
tangential component. The expected number of stars like FS2 is in good
agreement with our observation of the four other fast stars in NGC
2808.

Keeping in mind all the uncertainties that enter these estimates, we
consider both of these numbers consistent with the observation of one
star like FS1 and one like FS2.

%__________________________________________________________________

\subsection{The unbound case - Acceleration in the giant phase}
\label{sec:gienc}

The measured velocities for stars FS1 and FS2 are very close to the
cluster escape velocity. This makes it hard to establish whether the
two stars are bound or unbound to the cluster.  Here we consider the
case in which FS1 and/or FS2 received a recoil velocity during
a recent dynamical encounter large enough to unbind them from the
cluster.

We test the possibility that FS1 and FS2 were ejected from the cluster
core by a dynamical encounter by performing numerical three-body
scattering experiments similar to the ones described in Section
\ref{subsec:msenc}.  In this case, considering an ejection velocity of
$45\kmps$ and the current projected distance of $\sim 0.3\pc$ from the
cluster center, we can assume that the encounters that accelerated the
two stars took place when the stars were already in the giant
phase. This is an important assumption, and the one that distinguishes
the encounters described in this Section from those in
\ref{subsec:msenc}. Collisions in close encounters affect giant stars
much more than they affect main-sequence stars due to the larger radii,
and therefore larger cross-sections, of the giants.

\begin{table}
\caption{Dynamical encounters between a binary and a single star. FS1
  and FS2 indicate the giant stars under consideration, MS indicates a
  main-sequence star, and BH a stellar mass black hole. Outcomes of
  interest for the ejection of one of the giant stars include flybys
  (F), exchanges (E) and ionizations (I).}
\label{tab:encgi}
\centering     
\begin{tabular}{lll} 
\hline
\noalign{\smallskip}
Name & Encounter & Outcomes\\
\noalign{\smallskip}
\hline                        
\noalign{\smallskip}
G1 & (FS1, FS1) + FS1 & F, E, I \\
G2 & (FS2, FS2) + FS2 & F, E, I \\
M1 & (FS1, FS1) + MS  & E, I\\
M2 & (FS2, FS2) + MS  & E, I\\
M3 & (FS1, MS) + MS   & E, I\\
M4 & (FS2, MS) + MS   & E, I\\
B1 & (FS1, FS1) + BH  & E, I\\
B2 & (FS2, FS2) + BH  & E, I\\
\noalign{\smallskip}
\hline                             
\end{tabular}
\end{table}

We consider three main types of binary-single star encounters, which
are detailed in Table\,\ref{tab:encgi}: (1) encounters involving only
giant stars (models G1, G2), (2) encounters involving giant and
main-sequence stars (models M1, M2, M3, M4), and (3) encounters
between a binary of giant stars and a black hole (BH) (of both the
stellar and intermediate-mass kind). For each type of encounter listed
in Table\,\ref{tab:encgi}, we consider all outcomes resulting in the
ejection of (at least) one giant star, and record the ejection
velocity. Since we are interested in the ejection of giant stars with
large velocities, we neglect all cases of collisions.

Encounters of star FS1 and FS2 with other giants (case G1 and G2), or
main-sequence stars (case M1, M2, M3, M4) result in ejection
velocities always much lower than $45\kmps$, for any value of the
initial binary separation. It thus ruled out that any of the two high
velocity giants were ejected by encounters with stars of the same
mass, whether giants or main-sequence stars.

\begin{figure}
  \centering \includegraphics[angle=270,width=0.5\textwidth]{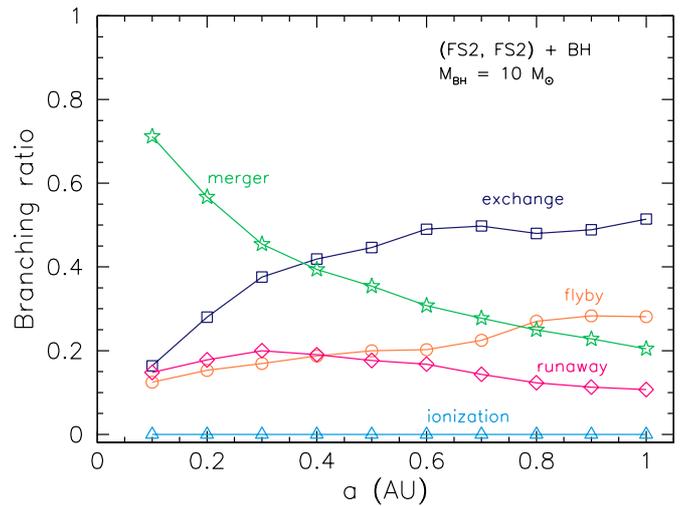}
  \caption{Branching ratios versus initial binary separation for the
    outcomes of encounters between a stellar binary and a black hole
    of $10\msun$ (model B2). The label ``runaway'' refers to
    encounters that produce an escaper with velocity larger than
    $45\kmps$.}
  \label{fig:branch}
\end{figure}

We also consider the case of encounters with black holes (BHs) of
different masses. Figure \ref{fig:branch} gives the branching ratios,
as a function of the initial binary separation, for model B2.  The
relative importance of mergers and exchanges depends sensitively on
the binary semi-major axis, with mergers occurring in more than 50\%
of the cases for $a \lap 0.2\au$.

\begin{figure*}
  \centering
  \includegraphics[angle=270, width=0.49\textwidth]{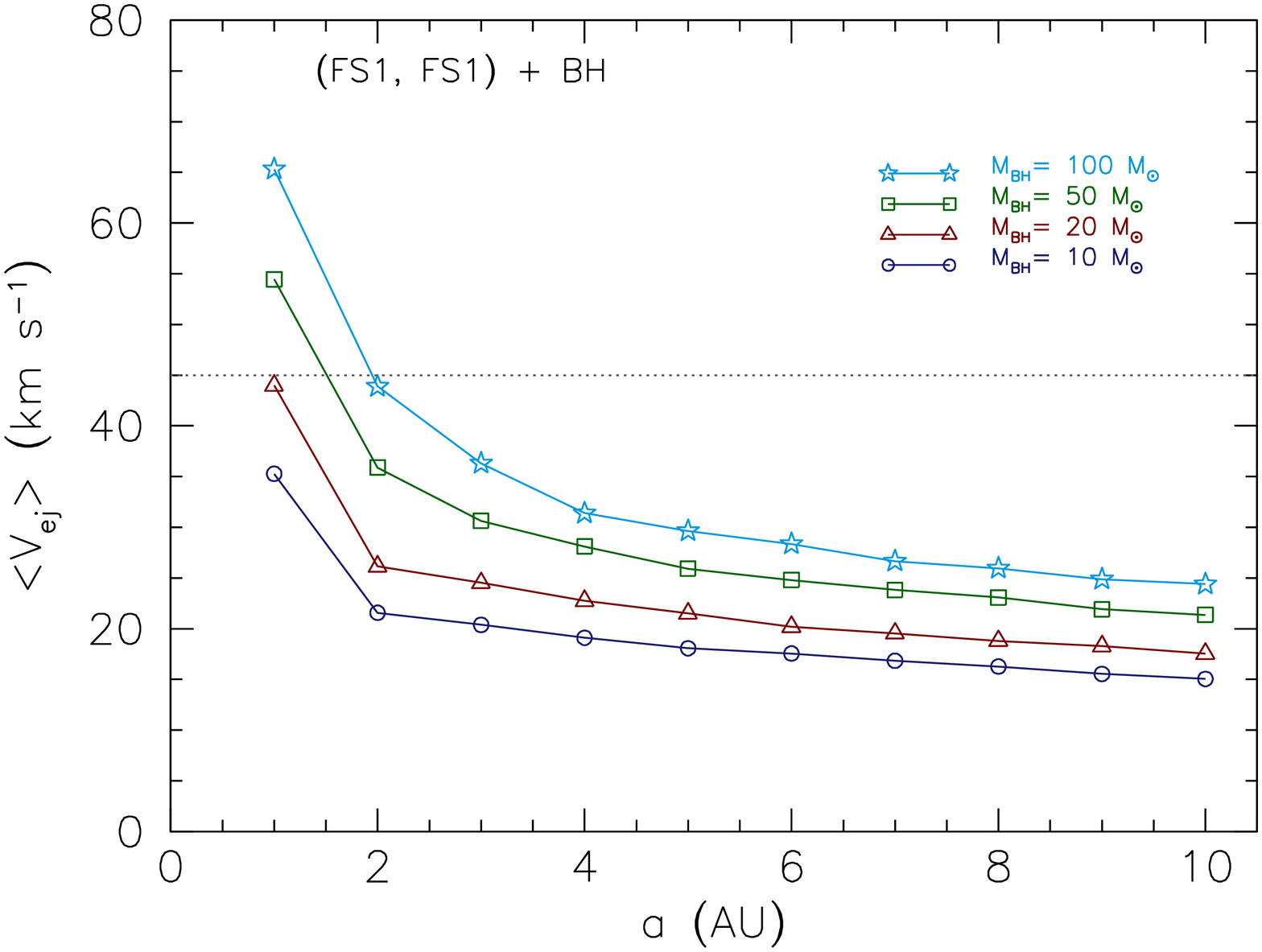}
  \includegraphics[angle=270, width=0.49\textwidth]{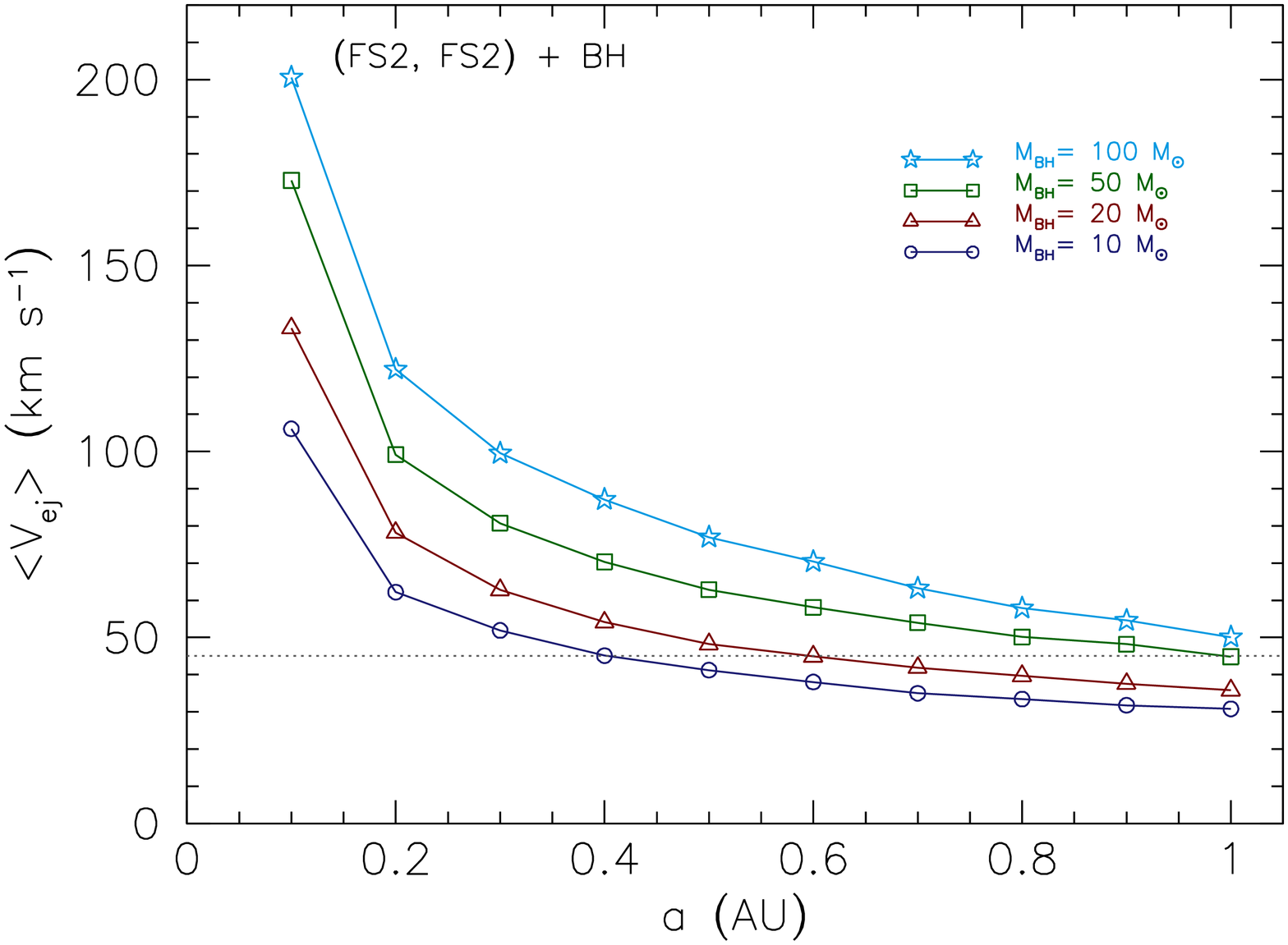}
  \caption{Average ejection velocity in runs B1 (left) and B2 (right)
    as a function of the initial binary semi-major axis. Different
    lines refer to different values of the BH mass, $M_{\rm BH} = 10,
    20, 50, 100\msun$, respectively, from bottom to top. Dotted lines
    indicate the required ejection velocity of $45\kmps$.}
  \label{fig:velo}
\end{figure*}

\begin{figure}
  \centering \includegraphics[angle=270,width=0.5\textwidth]{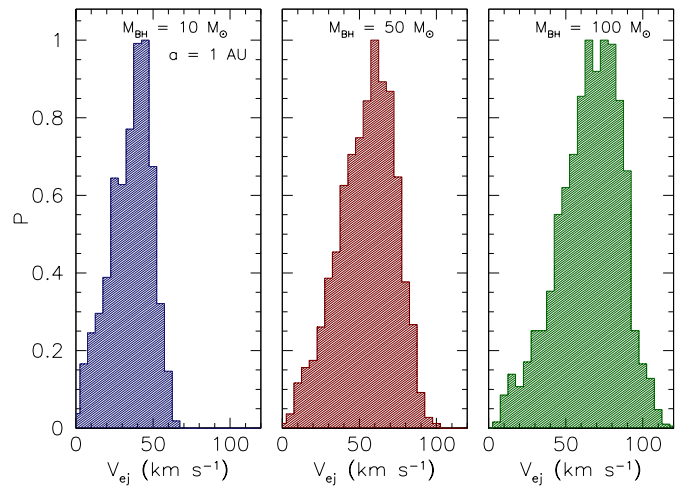}
  \caption{Distribution of ejection velocities in encounters between a
    binary of giant stars and a black hole (case B1). The three panels
    refer to BHs of different mass: $10, 50, 100\msun$, from left to
    right. The binary separation is set to $1\au$.}
  \label{fig:veld}
\end{figure}

Average ejection velocities as a function of the initial binary
semi-major axis are shown in Fig.\,\ref{fig:velo}, for models B1 and
B2, and for different values of the BH mass $M_{\rm BH} = 10, 20, 50,
100\msun$. Velocity distributions for case B1 are shown in
Fig. \ref{fig:veld} for BH masses of $10, 50, 100\msun$ and the
smallest value of the binary semi-major axis $a=1\au$. We find that
stellar-mass BHs are sufficient to accelerate star FS2 to velocities
of $\sim 45\kmps$ for any value of the binary separation, and
no BH of intermediate mass is required in the cluster to explain its
velocity. Average ejection velocities for star FS1 are somewhat
lower, due to the much larger size and therefore to the much larger
number of close encounters that end with a collision. However, if we
look at the distributions of ejection velocities for model B1, which
are shown in Fig. \ref{fig:veld} for the case $a = 1\au$, we find that
a $10\msun$ black hole is sufficient to produce ejection velocities of
$40-45\kmps$.

We compute the ejection rates for this ejection mechanism in the same
way as in Section \ref{subsec:msenc}. To obtain the final number of
ejected stars, however, the relaxation time in Equation \ref{eq:r}
needs to be replaced with the crossing time of the core. With a
typical velocity of $13\kmps$, the crossing time for a core diameter
of $1.4\pc$ is $T_{\rm cr} \sim 10^5 \yr$. This results in a final
number of events as listed in Table \ref{tab:num}. It is clear that
the higher the black hole mass, the more stars with velocities larger
than $45\kmps$ are produced. However, even a black hole of $10^3
\ \msun$ would not be enough to produce an acceptable number of stars
to explain the observations.

\begin{table}
\caption{Expected numbers of stars if accelerated by an encounter
  of a giant and a massive black hole.}
\label{tab:num}
\centering     
\begin{tabular}{lcccc} 
\hline
\noalign{\smallskip}
&\multicolumn{4}{c}{NUMBER OF EXPECTED STARS}\\
	 & $10 \ \msun$ & $20 \ \msun$ & $50 \ \msun$ & $100 \ \msun$ \\
\noalign{\smallskip}
\hline                        
\noalign{\smallskip}
FS1 & $3.17 \times 10^{-6}$ & $1.46 \times 10^{-5}$ & $3.38 \times 10^{-5}$  & $3.77 \times 10^{-4}$ \\
FS2 & $1.70 \times 10^{-5}$  & $2.21\times 10^{-5}$ & $3.04 \times 10^{-5}$  & $6.37 \times 10^{-5}$ \\
\noalign{\smallskip}
\hline                             
\end{tabular}
\end{table}

We conclude that a dynamical encounter between a binary of giant stars
and a stellar-mass black hole of $\gap 10 \msun$ can produce giant
stars with velocities $\sim 45\kmps$. However, the number of expected
fast stars produced by this mechanism is one to three orders of
magnitudes lower than our observations show. The reasons for this
discrepancy are, on the one hand, the lower ejection rates due to the
shorter timescales in which the stars will disappear from the field of
view and, on the other hand, the fact that giant stars are less likely
to achieve large velocities than main-sequence stars. For these
reasons, this scenario is unlikely to have accelerated the observed
runaways.

\section{Comparison with other globular clusters}\label{sec:others}

In Section \ref{intro} we mentioned the cases of M3 and 47 Tuc and
their high-velocity stars. \cite{meylan_1991} argued that the two fast
stars in 47 Tuc, which move with $3.6 \sigma_c$ and $4.8 \sigma_c$,
must have been recently accelerated by encounters with a giant star,
similar to our unbound case. The reason for this assumption is the
fact that the velocities of these stars ($\rm V_1 = -36.7$ \kms, $\rm
V_2 = 32.4$ \kms) are, within the errors, equal to the escape velocity
$\rm V_{esc} = 35.38$ \kms \citep{peterson_1975}. This is a very
similar situation to our case and therefore we would be more cautious
in claiming the stars are bound or unbound. The argument of
\cite{meylan_1991} that these stars could only recently have been
accelerated is questionable. Furthermore, even if the velocities of
the stars exceed the escape velocity by a few \kms, the probability
that the stars are able to actually escape the cluster immediately
remains low, since they are only able to escape by passing through
small apertures near the Lagrange points L1 and L2. This can result in
a significant delay of the escape of stars \citep{fukushige_2000}. For
this reason it is a valid hypothesis that these stars also got
accelerated in their main-sequence stage. Because the central
density\footnote{With the caveat that $\log(\rho_0)$ is measured with
  a different technique than our derived value and therefore lower
  than our actual central density for NGC 2808.}  $\log(\rho_0/\pcc) =
4.88$ and core relaxation time $\log(t_c/\yr) = 7.85$
\citep{harris_1996} of 47 Tuc are very similar to the values of NGC
2808, the expected numbers of fast stars will be close to our derived
numbers.

In the case of M3 the central velocity dispersion is lower than the
one of NGC 2808 by 50~\% ($\sigma_c = 4.9$ \kms). However, it is
important to note that the absolute velocity dispersion is not the key
factor, but rather its value relative to the velocities of the
high-velocity stars. \cite{gunn_1979} measured velocities of $17.0$
\kms and $-22.9$ \kms for two stars which are most likely cluster
members. Again, these velocities lie in the same range as the escape
velocity ($\rm V_{esc}=20.81$ \kms) but, as in the two previous cases,
do not exceed this limit by much. The only difference with NGC 2808 and
47 Tuc is the central density of $\log(\rho_0) = 3.57$. This would
lower othe expected number of high-velocity stars by an order of
magnitude. Since this would most strongly affect the prediction for
the unbound case, the most likely scenario for M3 is still that the
stars are bound and were accelerated by a dynamical encounter with a
massive object while they were in the main sequence phase.

The globular cluster NGC 6752 shows two millisecond pulsars
  outside the core: a recycled binary pulsar at more than three half-mass
  radii and a single pulsar at 1.4 half-mass radii. Such systems
  suggest the presence of a massive object in the center of the
  cluster. \citet{colpi2002,colpi2003}  show that the
  off-center location of the pulsars can be explained with scattering
  from a binary of stellar mass black holes or a single
  intermediate-mass black hole. This supports the idea that
  high-velocity stars in globular clusters may be produced by
  single or binary black holes.

%__________________________________________________________________

\section{Summary and conclusions}\label{sec:con}

We detect five high-velocity stars in the globular cluster NGC 2808
using integral field unit spectroscopy as well as Fabry-Perot data for
individual stars. All stars are $\sim 0.3 - 0.5\pc$ away from the
cluster center and their velocities correspond to $3.6 - 4.1
\sigma_c$, where $\sigma_c$ is the velocity dispersion in the
core. All stars are very likely to be cluster members due to their
positions in the CMD. Furthermore, we find that the proper motions of
the stars are in good agreement with the proper-motion distribution.

For the case of the two fast stars found in the integral-field unit,
we discuss various scenarios which could explain their peculiar
motions. By performing a binary synthesis and looking at HST far UV
and near UV images, we exclude the possibility of these stars being
close binaries. Furthermore, we discuss the possibility of
atmospherically acitve stars such as long period variables and
conclude that this scenario can also be excluded due to a lack of
spectroscopic evidence and general inconsistency with the observed
velocity range.

The measured velocities of the stars are close to the estimated
escape velocity from NGC 2808. For this reason, we consider two cases
of acceleration: The bound case, in which the stars were accelerated
while in the main-sequence stage; and the unbound case, in which the
acceleration must have taken place recently. We perform numerical
three-body scattering experiments for both cases. In the bound case we
also test the possibility that these stars belong to the tail of the
Maxwellian velocity distribution by means of Monte Carlo
simulations. In the case of three-body encounters of main-sequence
stars and compact objects (the bound case) we consider four types of
encounters: (1) encounters with a main-sequence star of the same mass,
(2) encounters with a $0.8\msun$ white-dwarf, (3) encounters with a
$1.4\msun$ neutron star and (4) encounters with a $\sim 10\msun$ black
hole. To obtain the expected numbers of ejected stars, we calculate
the event rates of each encounter producing a star with a velocity
larger than 45 \kms as a function of the binary separation. Combining
the rates with the log-normal orbital period distribution obtained by
\cite{duquennoy_1991}, we obtain the total event rate by integration.

Similarly, we test for the possibility of encounters with giant stars
(the unbound case). As an input, we take the estimated properties of
the two stars (radius and mass) and consider three main types of
binary-single star encounters: encounters between giants, encounters
of giant and main-sequence stars, and encounters including a black
hole. We find that all encounters including only stars result in much
lower escape velocities than the observed ones. Encounters between a
binary of giants and a black hole, however, result in velocity
distributions which peak at the observed velocities. Especially a
black hole of $M_{BH} \sim 10\msun$ and a binary separation of $a
= 1\au$ predicts velocities around 40 \kms for star FS1. For star FS2,
any separation produces large enough ejection velocities.

The scattering experiments show that no intermediate-mass black hole
is needed in order to accelerate the stars to their
velocities. Furthermore, only the encounters of main-sequence stars
and compact objects have rates high enough to explain the observed
stars.  This leads us to the conclusion that the stars are bound and
were accelerated by an encounter with a $\sim 10 \msun$ stellar black
hole. This cluster must thus have a high concentration of stellar-mass
black holes in its center in order to make such an encounter likely
enough to be observed. This is in agreement with the latest results on
the central kinematics of NGC 2808 where no intermediate-mass black
hole is detected \citet{nora11b}. Such a retention of stellar remnants
in evolving globular clusters is also in line with theoretical
expectations, which allow up to $30 \% - 50\%$ of the cluster mass to
be constituted by remnants \citep{baumgardt_2003, kruijssen_2009}.

We compare our results with the two other known cases in which high
velocity stars were detected in globular clusters, M3 and 47
Tuc. These cases are very similar to NGC 2808 and might therefore be
explained with the same arguments. The high-velocity stars in these
clusters are most probably products of encounters of main-sequence
binaries with a massive black hole. For further analysis, N-body
simulations could help verifying these results by running specific simulations
dedicated to the statistics of fast stars and their origin. When
combined with an analysis as carried out in this paper, this provides
a general prediction for the origin of high-velocity stars in globular
clusters.

%__________________________________________________________________

\begin{acknowledgements}
We thank Thomas Lebzelter, Mathieu Servillat and Dietrich Baade for
sharing their knowledge and expertise. Also, special thanks go to Jay
Anderson and his group for providing us with the proper motions and
for the strong support of this project.  This research was supported
by the DFG cluster of excellence Origin and Structure of the Universe
(www.universe-cluster.de). H.B. acknowledges support from the
Australian Research Council through Future Fellowship grant
FT0991052. We thank the referee, Steinn Sigurdsson, for constructive comments
that helped improve the paper.

\end{acknowledgements}

%Jay, Rodolfo, Dietrich, Lebzelter

\bibliographystyle{aa}
\bibliography{ref}

\end{document}